\documentclass[12pt]{article}
\usepackage{amssymb}
\usepackage{amsmath}

\def\fun#1#2{\lower3.6pt\vbox{\baselineskip0pt\lineskip.9pt
\ialign{$\mathsurround=0pt#1\hfil##\hfil$\crcr#2\crcr\sim\crcr}}}
\newcommand{\beq}{\begin{equation}}
\newcommand{\eeq}{\end{equation}}

\usepackage{epsfig}
\newcommand{\be}{\begin{equation}}
\newcommand{\ee}{\end{equation}}
\newcommand{\bea}{\begin{eqnarray}}
\newcommand{\eea}{\end{eqnarray}}

\begin{document}
\title{Impact of the ionization of the atomic shell on the lifetime of the $^{229m}$Th isomer.}
\author{F. F. Karpeshin$^1$\footnote{e-mail: fkarpeshin@gmail.com}, M.B.Trzhaskovskaya$^2$, L.F.Vitushkin$^1$ }

\maketitle
\begin{center}
$^1$D.I.Mendeleyev Institute for Metrology, St.-Petersburg, Russia  \\
$^2$PNPI, Kurchatov Center, St.-Petersburg, Russia  
\end{center}

    \begin{abstract}
Contemporary data are analyzed concerning the half-lives of the $^{229m}$Th isomer in neutral atoms and various ions. It is explicitly shown that the isomer  lifetime may strongly depend on the plain environmental physical conditions like  pressure and temperature. 
Calculation is performed on the united platform of interplay of traditional and subthreshold resonance conversion. General agreement with experiment is obtained in the cases of Th$^I$ and Th$^{III}$, a prediction is made concerning Th$^{IV}$. Disagreement with the  experiment in the case of Th$^{II}$ may be a consequence of a random very sharp resonance with an electronic level. Account of the influence of the environment on the decay rate helps one to loose the resonance condition, which may lead to agreement with experiment.  This will mean the first experimental  observation of strong dependence of the nuclear decay rate on the environment. 
\end{abstract}

\bigskip

\clearpage

\section{Introduction}

The idea of the combined atomic-nuclear transitions experienced  a period of strong development during the last time. Making use of the resonance properties of the electron shell opens the way of manipulating the nuclei.  A possible practical realization of this way is the creation of the optical-nuclear clock based on a few-eV nuclear  isomer of $^{229}$Th \cite{pike,clok2}. For this purpose, more detailed information concerning the isomer properties is still needed,  including the exact value of the isomer energy and its half-life time.

      Indirect evidence of the presence of the isomeric level is known for decades. However, nobody could detect the isomer or its decay directly. Only recently, its decay through internal conversion (IC) was finally discovered \cite{lars}. Nevertheless, information about characteristic properties of the isomer, including its precise energy and lifetime, remains extremely scarce.   Thus, the estimate of its energy varies in time. An energy of 3.5 eV was considered for a long time \cite{reih}. Sometimes, a value of 5.5 eV was also used \cite{gimar}. Most recent measurements resulted in a higher value of 7.6$\pm$1 eV \cite{beck}. However, other values are also checked and cannot be excluded (e.g., \cite{lars,larsnew} and refs. cited therein). For the present purposes, we are oriented to this value as the latest data.

Moreover, the isomer half-life was measured for the first time. Its value of 7 $\mu$s was obtained in neutral atoms \cite{larsnew}, in coincidence with the theoretical estimate \cite{prc}. In neutral atoms, the isomer energy is  higher than the ionization potential $I_a$ = 6.3067$_3$ eV \cite{nist}. Therefore, decay occurs via IC. The decay width is described by 
\be
\Gamma=(1+\alpha(M1))\Gamma_\gamma^{(n)}    \,, \label{Gc}
\ee with $\alpha(M1)$ being the internal conversion coefficient (ICC).
Ground state electronic configuration is $(7s)^2(6d_{3/2})^2$. With the calculated ICC value in the 7s electronic shell,   $\alpha(M1) = 1.1\times 10^9$ \cite{prc}, Eq. (\ref{Gc}) allows one to conclude on the radiative nuclear half-life to be $T_{1/2} \approx$ 2 hours.  

It is didactic to trace how the estimation of the lifetime depends on the energy of the isomer.
Let us consider Fig. 2 in \cite{prc}. The ICC is exactly inversely proportional to $\omega^{-3}$, where $\omega$ is the transition energy. On the other hand, the nuclear radiative width $\Gamma_\gamma^{(n)}$ is proportional to $\omega^3$. Thus the expected lifetime, which is inversely proportional to  ICC times $\Gamma_\gamma^{(n)}$, remains constant. A change in the lifetime could occur when switching on the next $6s$ shell to internal conversion, but this would occur only at 37 eV. Currently such energies are not under discussion.

       In the ionized atoms of $^{229}$Th, IC becomes energetically closed. However, its mechanism remains effective in the form of bound internal conversion (BIC), also called resonance conversion because of its resonance character. owever, HHHHH 
Let us consider this process in finer detail.

\section{Results in the case of BIC}

In the case of singly charged ions, the ionization potential is $I_a$ = 12.10$_{20}$ eV \cite{nist}, and the IC channel is energetically closed. However, deexcitation occurs mainly through many electronic bridges. For the first time this was shown in \cite{kabzon} for the 76-eV $^{235}$U isomer, and in  \cite{antib} in the case of neutral atoms of $^{229}$Th, under assumption of an isomer energy of 3.5 eV.  Calculations for singly charged ions of $^{229}$Th were performed in ref. \cite{PL2}.  More detailed calculations for the neutral atoms of $^{229}$Th, taking into account mixing of the electronic configurations, were performed in refs. \cite{prc,yaf}.  The results show that the main contribution comes from a few electronic transitions, in spite of the high fragmentation of the single-electron levels. For this reason, BIC remains to be a resonance process in its character, which does not exclude fluctuations of the lifetime depending on the accidental match of the energies of the nuclear and one of such a strong electron transition. With      a certain set of the basis electron configurations,  such a resonance enhancement was noted in \cite{prc,yaf}. As a result, it shortened the calculated lifetime by an order of magnitude.

      In Ref. \cite{atta}, the electronic bridges were considered for the 35-keV $M1$ transition in $^{125}$Te as the extension of traditional IC to the threshold case. The process received the name of BIC. Such an approach allows one to apply a conventional physical model and methods of calculations. The traditional ICC, $\alpha(\tau,L)$, go over the analogues $\alpha_d(\tau,L)$, which are obtained by mere replacement of the conversion electron wavefunctions in the continuum by those from the discrete spectrum.  They acquire dimension of energy as a result of such a replacement. To obtain the dimensionless resonance conversion factor $R$, one has to multiply this analogue by the resonance  Breit---Wigner factor. Then we come to a conventional expression, similar to (\ref{Gc}):
\be
\Gamma=(1+R) \Gamma_\gamma^{(n)}	\label{lt}
\ee   
where, in turn,  the discrete conversion factor $R$ is expressed in terms of the analogue of ICC, $\alpha_d$:
\be
R = \sum_i\frac{\alpha_d^{(i)}(M1) \Gamma_t^{(i)} /2\pi} {(\Delta^{(i)})^2+(\Gamma_t^{(i)}/2)^2}  \label{Rg}
\ee 
with
\be
\Gamma_t=\Gamma_\gamma^{(n)}+\Gamma_a\approx \Gamma_a
\ee
being the total width of the nuclear and atomic decays, and
\be
\Delta^{(i)} = \omega_n-\omega_a^{(i)}\,.
\ee --- the defect of the resonance. Summation in (\ref{Rg}) is over all the intermediate states. 

     Calculations were made within the Dirac-Fock method, taking into account the interaction of the electronic configurations. Simlar to IC in neutral atoms, it is the $7s-8s$ transition which predetermines the rate of the RC transition.

      Concerning the electron configuration of the ground state, different information can be noted in the literature. One possibility is
      \be
      (7s)^26d_{3/2} \label{7s}  \ee 
(e.g., Ref. \cite{nist}), the other is
      \be
      7s(6d_{3/2})^2 \label{6d} \ee   (e.g., Ref. \cite{handbook}). Our calculation shows that these two single-electron states are practically degenerate. For this reason, their  linear combination with comparable weights may be used in practice. Our calculation shows that the ground state is $7s(6d_{3/2})^2$. 

      Eq. (\ref{Rg}) deserves the following comment. It is conventionally accepted that the nuclear properties, specifically the radioactive decay constant, are essentially independent of the physical environment. At most, variations of the nuclear lifetimes are mentioned depending on the chemical emvironment. They were found not to exceed the level of  ten percent. The variations may arise due to change of the IC rate with the variation of the population of the upper electronic shells. Such a stability against the environmental medium  underlies the idea of the nuclear clock. 
To a certain sense,  somewhat aside is a number of papers, where it was shown that the nuclear processes may be strongly affected buy means of laser radiation, using electron shells as resonators \cite{kabzon,PL1,canad,Chin,Hf09,echa,book}. Specifically, it was predicted  that the decay of the $^{229}$Th isomer  would be by $\sim$700 times faster in the hydrogen-like ions \cite{zylic}. A similar effect was also predicted in $^{169}$Yb \cite{YbZETP}. But in fact, eq. (\ref{Rg}) says much more. Namely, as one can see from eqs. (\ref{lt}) -- (\ref{Rg}), in the case of BIC the atomic width $\Gamma_a$ enters the expression for the nuclear decay width directly as a factor, on the equal footing  with the radiative nuclear width and the discrete analogue of  ICC $\alpha_d$. That is, the nuclear decay width turns out to be directly proportional to the atomic width $\Gamma_a$, which in turn is determined by completely different factors, including  such as mere temperature and pressure of the physical environment. 

	The consecutive calculation of influence of the electronic bridges on the lifetime was undertaken in ref. \cite{prc,yaf} for the neutral Th atoms, under assumption of the isomer energy of 3.5 eV. The calculations were performed within the Dirac-Fock method, taking into account electron configuration mixing. The resulting value of $R\approx 600$ was argued. On a twist of fate, the resonance energy of the electronic bridge in Th$^I$ is approximately 3.5 eV, which just coincided with the energy of the isomer of 3.5 eV assumed therein \cite{prc,yaf}. In Th$^{II}$, the resonance energy, according to our calculations, is 7.6 eV, which is again in resonance with the presently adopted nuclear energy. Therefore, allowing for the aforementioned holding of the IC and RC rates with respect to the transition energy, the results of \cite{yaf,prc} should hold for this ion. Taking into account that there is only one $7s$ electron in the singly charged ions, as distinct from the neutral atoms,  the value of $R$ may be diminished by a factor of two. Thus we arrive at the value of    $R^{II}\approx$ 300. 
On the other hand, the $R$ value can be estimated directly by means of eq. (\ref{Rg}). Our calculations allow us to expect the presence of several  resonance states correlating with the single-electron $8s$ state in the range at $\sim$ 8 eV.  
There are, however, no famous $8s$ states at this energy, however, that would allow to extract a proper $\Delta$ value. \emph{Ab initio},  we can suppose it to be of about $\Delta\approx$ 1 eV. Experience of paper \cite{prc} teaches that due to configuration interaction, the resulting $R$ factor is not sensitive to the $\Delta$ value, if not an accidental very sharp resonance occurs. 

	With the results \cite{npa} for the $\alpha_d(M1)$, atomic widths and energies, we thus  arrive at a close value of $R\approx$ 100 \cite{izmt}. Therefore, with the accuracy which is quite enough for the present purposes we can surely rely on  an estimate of $R\approx$ 100 -- 300. We will use the latter value. The half-life time calculated by means of (\ref{lt}) turns out to be about 25  s then. Were configuration (\ref{7s}) assumed,  the half-life time would become twice shorter. In the case of linear combination, the lifetimes should be somewhat within this interval.

      From these estimations, one can conclude on a strong dependence of the Th$^{II}$ lifetime on the electron configuration.  Surprisingly, the calculated value appears to be  in contradiction with experimental data \cite{larsnew}. The latter show a value which is shorter than of about 10 ms. That is at least by three orders of magnitude less than the theoretical estimate. In search for probable reasons for such contradiction, let us allow for what is said above about the 	influence of the environment on the nuclear lifetime in the case of BIC. Then one order can be explained as a consequence of the collisional broadening of the atomic lines. 
An unexplained difference therefore remains within about two orders of magnitude. This may be interpreted as a consequence of a randomly close coincidence with a meaningful electron level. Such an example was considered in ref. \cite{prc}, when a random coincidence within 0.03 eV led to the change of the $R$ factor by an order of magnitude. At the samer time, in the present case the coincidence should be at the level of 0.01 eV, which seems to be one more caprice of fate. 

       Another possibility is to assume that the isomer energy is $\omega_n \gtrsim I_a$ =12.1 eV. In this case, the half-life time is again determined by IC, which  infers its value of $\sim$10 $\mu$s.
Further experiments should shed light on the question.

      In the case of Th$^{III}$, the calculated ground state configuration is $6d_{3/2}5f_{5/2}$. This is in agreement with available results from the literature \cite{nist,austral}.  Therefore, the electronic bridge through the transition $7s-8s$ is ruled out in the main approximation. According to what is said above, it may occur due to admixture of the $7s$, $8s$ and other $s$ states to the main configuration in the ground state.  Therefore,  the mixing coefficients comprise amplitudes of tenths at most, and the related $R$ values are suppressed by a factor of 10 at least. This
reduces the expected  $R$ value in comparison with the previous case of the single ions, to the values within   $R^{III}\lesssim$ 10. This means that the expected half-life becomes a step closer to that as due to a pure radiative deexcitation: 12 minutes $\lesssim T_{1/2} \lesssim$ 2 hours. This is in accordance with the observed value \cite{lars} of $T_{1/2} >$ 100 s. This allowed to conclude that the energy of the isomer is less than $\omega_n \lesssim I_a$ = 18.3 eV for these ions \cite{lars}.

      In the three-fold ions of Th$^{IV}$, the calculated ground state is $5f_{5/2}$. By analogy with the previous case, its expected lifetime should be close to that in the bare nuclei, that is about 2 hours. 

      \section{Conclusion}

We performed systematic analysis of the calculated lifetimes of the $^{229m}$Th isomer in ions of different degree of ionization, from neutral atoms to three-fold ions. Consideration is fulfilled on the unified platform of  interplay of traditional IC with subthreshold RC,  embodied by electronic bridges. Such a unified approach points clearly that the lifetime is not critical to exact values of the isomer energy, when the latter is within the domains above or lower than the potential of ionization $I_a$ of the atoms. It is mainly defined by the parameter of interrelation between the isomer energy and the ionization potential. The other parameter which defines the lifetime in the case of subthreshold isomer energy in  single ions of Th$^{II}$ is the relation between the weights of the ground electronic configurations (\ref{7s}) and (\ref{6d}).  Our calculation results in the ground  configuration  (\ref{6d}). This leads to the half-life  of 165 s, which would gradually shorten down to 80 s with switching on the other component  (\ref{7s}).

      The results are compared to available experimental data for neutral atoms, one- and
two-fold ions. First, we note the agreement which is achieved in the case of neutral atoms and two-fold ions. The former  says that the isomer energy exceeds 6.1 eV --- the ionization potential. It agrees with the most popular accepted value of 7.6 eV as the isomer energy.  Moreover, the agreement also confirms estimation of the nuclear radiative lifetime as approximately five hours for this energy.  Furthermore, assuming this energy value, the calculated lifetime in the two-fold ions also turns out to be in agreement with the experiment. In this case, the lifetime should be approximately as in the bare nuclei times a factor, which may vary from 1 to 10. This substantiates the general concept in frame of which we consider subthreshold resonance conversion, realized through the electronic bridges.

    Bearing this model in mind, all the more intriguing looks the disagreement with experimental data obtained for the single ions of Th$^{II}$. Experiment shows a value by a factor of  at least 100 times shorter. This may be explained in two ways. Either there is a random match of the nuclear energy with some meaningful electron transition energy better than within 0.01 eV.  This would be one more remarkable coincidence, along with the uniquely low energy.  And one should  bear in mind that the experimental value is an upper bound, the real half-life time may be still much  shorter.   That might also be understood if the isomer energy were somewhat higher, within 12.1 to 18.3 eV. In this case, the lifetime would be determined again by internal conversion, and therefore, it would be at the level of microseconds. At the same time, this possibility would not essentially affect the cases of Th$^I$ and Th$^{III}$. Further experimental  research is needed for clarification of this fact.

    Prediction is made concerning the half-life of the three-fold ions, which should be practically as in the bare nuclei --- of approximately five hours. We note a beautiful analogy between our model of variation of the half-lives with ionization, as obtained above, with that for 45 to 48 fold ions of  $^{125}$Te \cite{atta}. In both cases, the lifetime regularly changes as usual IC gives drive to subthreshold resonance conversion, which in turn fades away with further ionization.

\bigskip
The authors would like to acknowledge very helpful and highly heuristic discussions with L. Wense, which induced  the writing of this work.


\newpage
\begin{thebibliography}{99}\fussy


\bibitem{pike} {\it E. Peik and Chr. Tamm}, Europhys. Lett., {\bf 61},  181 (2003).

\bibitem{clok2} Campbell, C. J. et al. Phys. Rev. Lett. 108, 120802 (2012).

\bibitem{lars} L. Wense, B. Seiferle, M. Laatiaoui {\it et al.}, Nature, {\bf 533}, 47 (2016).

\bibitem{reih}  {\it R.G.Helmer and C.W.Reich}, Phys. Rev. {\bf C49}, 1845 (1994).

\bibitem{gimar} {\it Z. O. Guimaraes-Filho and O. Helene}, Phys. Rev. {\bf C71}, 044303 (2005).

\bibitem{beck} {\it B. R. Beck et al.,} Phys. Rev. Lett., {\bf 98},  142501 (2007).

\bibitem{larsnew} B. Seiferle {\it et al.}, Phys. Rev. Lett., {\it at press}.

\bibitem{prc} F. F. Karpeshin, M. B. Trzhaskovskaya, Phys. Rev. C, vol. 76, p. 054313 (2007).

\bibitem{nist} Kramida, A., Ralchenko, Yu., Reader, J., and NIST ASD Team (2015). NIST Atomic Spectra Database (ver. 5.3), [Online]. Available: http://physics.nist.gov/asd [2016, December 3]. National Institute of Standards and Technology, Gaithersburg, MD.

\bibitem{kabzon} B.A.Zon,  F.F.Karpeshin, Sov. Phys. --- JETP, vol. 70, p. 224, 1990.

\bibitem{antib} F.F.Karpeshin,  I.M.Band and M.B.Trzhaskovskaya, in:
Nuclear Shapes and Nuclear Structure at Low Excitation Energies. Antibes (France) 20 -- 25 June, 1994. Abstracts of the Contributed Papers, p. 50; Proceedings of the International Conference.

\bibitem{PL2}  F.F.Karpeshin, I.M.Band, M.B.Trzhaskovskaya  and     M.A.Listengarten,  Phys. Lett. {\bf B372} (1996) 1.

\bibitem{yaf}   F. F. Karpeshin, M. B. Trzhaskovskaya, Yad. Fiz. {\bf  78}, 765 (2015).  ({\it In Russian. Engl transl.:})  Phys. At. Nucl. {\bf  78}, 715 (2015).

 \bibitem{atta} F.\,F.\ Karpeshin, M. R. Harston, F. Attallah, J. F. Chemin,
     J. N. Scheurer, I. M. Band and M.\,B.\ Trzhaskovskaya, Phys. Rev. C
{\bf 53} (1996) 1640.

\bibitem{handbook} J. E. Sansonetti and W. C. Martin, Handbook of Basic Atomic Spectroscopic Data.   J. Phys. Chem. Data, {\bf 34}, 1559 (2005).

\bibitem{PL1} F.F.Karpeshin,  I.M.Band, M.B.Trzhaskovskaya and B.A.Zon,
Phys. Lett. {\bf 282B}, 267, 1992.

\bibitem{canad} F.F.Karpeshin,   M.A.Listengarten,  B.A.Zon, I.M.Band and M.B.Trzhaskovskaya,  Can. J. Phys., {\bf70}, 623 (1992).

\bibitem{Chin} F.F. Karpeshin, Zhang Jing-Bo and Zhang Wei-Ning,
    Chinese Physics Letters, {\bf 23} (2006) 2391.

\bibitem{Hf09} F.F. Karpeshin, M.B. Trzhaskovskaya and J. Zhang, Eur. Phys. J.  {\bf A 39}, 341 (2009).


\bibitem{echa} F. F. Karpeshin, Particles and Nuclei, {\bf  37}, 522 (2006).

\bibitem{book} F. F. Karpeshin, Prompt Fission in Muonic Atoms and Resonance Conversion. Saint-Petersburg: Nauka, 2006.

\bibitem{zylic} F. F. Karpeshin,  S. Wycech,  I. M. Band, 
       M.\,B.\ Trzhaskovskaya, M. Pf\"utzner and J.\.Zylicz, Phys. Rev. {\bf C57}, 3085, 1998.

\bibitem{YbZETP}  F.\,F.\ Karpeshin, M.B.Trzhaskovskaya and
    Yu.P.Gangrsky, Zh. Eksp. i Teor. Fiz., 2004, {\bf 126}, 323
    {\it  Engl. transl.} JETP, {\bf 99}, 286 (2004).

\bibitem{npa} F.F.Karpeshin,  I.M.Band and M.B.Trzhaskovskaya,  Nucl.
    Phys. {\bf A654}, 579 (1999).

\bibitem {izmt} F.F.Karpeshin,  M.B.Trzhaskovskaya,  Measurement Techniques, No. 7, 2016.

\bibitem{austral} E. Bi{\'e}mont, P. Palmeri, P. Quinet, Z. G. Zhang, And S. Svanberg, Astrophys. J., {\bf 567}, 1276 (2002).



\end{thebibliography}
\end{document}